\documentclass{aa}
\usepackage{psfig} 
\usepackage{graphicx}
\title{A quantitative analysis of OCN$^{-}$ formation in interstellar ice analogs}
\author{F.A. van Broekhuizen\inst{1}, J.V. Keane\inst{2}, and W.A. Schutte\inst{1}}
\offprints{F.A. van Broekhuizen,\\e-mail: fvb$@$strw.leidenuniv.nl}
\institute{$^1$ Raymond and Beverly Sackler Laboratory for Astrophysics, Leiden
Observatory, P.O. Box 9513, 2300 RA Leiden, The Netherlands\\
$^2$ NASA-Ames Research Center, Mail Stop 245-3, Moffett Field, CA 94035, USA}
\date{18 June 2003}
\authorrunning{F.A. van Broekhuizen et al.}
\titlerunning{A Quantitative Ananlysis..}
\begin{document}
\abstract{The 4.62\,$\mu$m absorption band, observed along the line-of-sight towards various young stellar objects, is generally used as a qualitative indicator for energetic processing of interstellar ice mantles. This interpretation is based on the excellent fit with OCN$^-$, which is readily formed by ultraviolet (UV) or ion-irradiation of ices containing H$_{2}$O, CO and NH$_{3}$. However, the assignment requires both qualitative and quantitative agreement in terms of the efficiency of formation as well as the formation of additional products. Here, we present the first quantitative results on the efficiency of laboratory formation of OCN$^{-}$ from ices composed of different combinations of H$_{2}$O, CO, CH$_{3}$OH, HNCO and NH$_{3}$ by UV- and thermally-mediated solid state chemistry. Our results show large implications for the use of the 4.62\,$\mu$m feature as a diagnostic for energetic ice-processing. UV-mediated formation of OCN$^{-}$ from H$_{2}$O/CO/NH$_{3}$ ice matrices falls short in reproducing the highest observed interstellar abundances. In this case, at most 2.7\% OCN$^{-}$ is formed with respect to H$_{2}$O under conditions that no longer apply to a molecular cloud environment. On the other hand, photoprocessing and in particular thermal processing of solid HNCO in the presence of NH$_{3}$ are very efficient OCN$^-$ formation mechanisms, converting 60\%--85\% and $\sim$100\%, respectively of the original HNCO. We propose that OCN$^-$ is most likely formed thermally from HNCO given the ease and efficiency of this mechanism. Upper limits on solid HNCO and the inferred interstellar ice temperatures are in agreement with this scenario.   
}
\maketitle 
\section{Introduction}
\indent The chemical composition of interstellar ices acts as a tracer of the chemical and physical history of molecular clouds. Based on spectral absorption features in the 2--16\,$\mu$m region, the ices are found to consist of simple neutral molecules. In addition, observational evidence now exists for the presence of ions such as OCN$^-$, HCOO$^-$ and NH$_4^+$ (Grim et al. 1989; Allamandola et al. 1999; Novozamsky et al. 2001; Schutte \& Khanna 2003). The formation mechanisms underlying their presence, however, are still uncertain for most species. Understanding these processes is of key importance to improve our knowledge of the ice history and the evolution of molecular clouds.\\
\indent The 4.62\,$\mu$m (2165 cm$^{-1}$) laboratory feature of OCN$^-$ matches well an interstellar absorption band, observed towards various young stellar objects (YSO's). Ever since its first detection in the infrared spectrum of W\,33\,A by Soifer et al. (1979), its assignment has remained controversial. First laboratory studies by Moore et al. (1983) reproduced the interstellar feature by proton irradiation of interstellar ice analogs composed of H$_{2}$O, CO and NH$_{3}$. Subsequently, Lacy et al. (1984) produced a similar feature by Vacuum ultraviolet (VUV) photoprocessing of CO/NH$_{3}$ ice. This `XCN' feature was assigned to OCN$^{-}$ by Grim \& Greenberg (1987), and Schutte \& Greenberg (1997) strengthened the hypothesis by looking at the band shift using different isotopes. In addition, Demyk et al. (1998) showed that OCN$^{-}$ could also form from HNCO by simple acid-base chemistry with NH$_3$, serving as a proton acceptor. This process has recently been studied in more detail by Raunier et al. (2003a, b). Although HNCO has not been identified directly in interstellar ices, it is observed in `hot cores' surrounding massive protostars and is a likely precursor of OCN$^-$. In spite of the strong case for assigning the interstellar 4.62\,$\mu$m feature to OCN$^{-}$, many other carriers were proposed. A summary is given by Pendleton et al. (1999). Some carriers, like silanes, thiocyanates, isothiocyanates and ketenes, cannot be excluded but to date none are serious candidates.\\
\indent The ease with which the interstellar 4.62 $\mu$m feature can be reproduced in the laboratory by the energetic processing of simple ices led to the direct association of this band in interstellar spectra with the presence of energetic processes. This is, however, a tentative assumption. A careful interpretation of interstellar data requires a thorough quantitative study of the efficiency of the OCN$^{-}$ formation together with an exhaustive investigation of the formation of additional products. Moreover, renewed interest has stemmed from the possibility that CN-bearing species may get included in protoplanetary discs to become an important component of prebiotic matter. This adds to the importance in the quest to understand the underlying formation mechanism of one of its simplest compounds, OCN$^{-}$ (Tegler et al. 1995; Pendleton et al. 1999; Whittet et al. 2001).  \\ 
\indent This study will look in detail at the formation efficiency of OCN$^{-}$ by UV- and thermal processing of interstellar ice analogs and the evolution of additional products under laboratory conditions. The outline of the paper is as follows. Sect. 2 describes the experimental procedure, leading to the results in Sect. 3. The initial precursor abundance and the influence of photon-energy distribution and ice-thickness on the efficiency of OCN$^-$ formation are addressed. Also alternative C- or N-sources to form OCN$^-$ other than CO or NH$_3$ are considered, namely, H$_2$CO, CH$_4$ and N$_2$. Special attention is paid to the formation of OCN$^{-}$ from HNCO or CH$_3$OH. Sect. 4 discusses the astrophysical implications of the results in relation to the 4.62\,$\mu$m `XCN' feature observed in ices towards various YSO's and aims to provide a more solid base for the use of this feature as a potential diagnostic of energetic processing of ices in space. Finally in Sect. 5 our conclusions are summarised.
\section{Experimental methods}
All experiments were performed under high vacuum ($10^{-8}$ mbar) conditions using Fourier Transform Infrared spectroscopy (FTIR) over the 4000--400 cm$^{-1}$ spectral range at 2 cm$^{-1}$ resolution. Ice matrices were grown at 15 K on a CsI window following the general procedure as described by Gerakines et al. (1995, 1996). To obtain a particular ice composition, constituent gases were mixed in a glass vacuum manifold (typical base pressure 2$\times10^{-4}$ mbar) to their initial relative ice abundances. Unless stated otherwise, the ices were grown as heterogeneous mixtures.\\
\indent Cyanic acid, HNCO, was produced as described by Novozamsky et al. (2001). In addition to their protocol, the collection of volatile HNCO was started at a cracking temperature of 343 K using a liquid nitrogen cool-trap and continued for 5 min. after completion of cracking. Infrared analysis of the volatile products at 216 K showed HNCO and CO$_{2}$ together with some very weak unidentified features. HNCO was purified by vacuum freeze-thaw using a n-octane and an isopentane slush at 216 K and 154 K, respectively. The purity of the HNCO was always $>99.5\%$ relative to CO$_{2}$. The contamination level of the residual products is probably less than 50\% of the CO$_{2}$ impurity and is thus thought to have a minor influence on the experimental results of HNCO in astrophysically relevant ice matrices. Therefore no further purification was applied.\\
\indent The thermal formation of OCN$^{-}$ was studied by raising the temperature of the ice matrix in steps of 5--20 K. The actual warmup was fast relative to the specific time of the typical experiment such that no reaction rate constants could be determined.\\ 
\indent Photoprocessing of optically thin ices used a microwave H$_{2}$-discharge lamp equipped with a MgF$_{2}$ window (Gerakines et al. 1996) to obtain the standard 'hard' UV with an overall UV flux of $\sim$$5\times10^{14}$ photons cm$^{-2}$ s$^{-1}$ (E$_{\rm{photon}} \geq{6}$ eV). These conditions simulate best clouds located near hot stars ($\sim$20000 K) or cosmic-ray induced UV photons inside a molecular cloud. A less energetic `soft' UV source with a $\sim$$5\times$ lower UV flux was produced by using a fused silica window (Janos technology ICC.). Fused silica becomes opaque below 140 nm. The transmitted photons are therefore lower in energy, typically E$_{\rm{photon}} \leq{8.8}$ eV. Astrophysically, this less energetic photon source can be viewed as a rough simulation of the UV impinging on ices that reside in more shielded regions around YSO's where the higher energy photons have been absorbed by dust. Both hard and soft energy spectra are extensively described by Mu\~noz-Caro \& Schutte (2003).\\
\indent Any astrophysical implications, based on efficiencies found in the laboratory must be treated with care. In the laboratory the ice is grown on a CsI-substrate. CsI can give rise to surface directed chemistry which can be different from interstellar grains. This effect is assumed to play a minor role within ice-thicknesses studied here. Also, differences in surface area introduce changes such as UV induced heating effects. Local heating on grains predominantly influences the very small ones that account for only a minor part of the total population and are therefore assumed to introduce only a small error on the total results. One other important point to consider is the total UV-flux of a molecular cloud, translating to a total UV fluence seen by one individual grain. The laboratory UV-flux is kept such that it predominantly induces single-photon processes. Still the flux in the laboratory is about 10 orders of magnitude more intense than that inside a molecular cloud. UV-photons induce secondary electrons and can form metastable species that can react on different time scales than ion-molecule or radical-molecule reactions. These processes, when present, could lead to discrepancies between observations and laboratory results.
\section{Results}
\subsection{Band strengths}
The integrated band strength, $A$, was determined for the relevant HNCO, OCN$^{-}$, and NH$_{4}^{+}$ features and was derived from the optical depth, $\tau_{\nu}$, at a frequency $\nu$ by means of the following equations where $N$ is the total amount of molecules.
\begin{equation}
{A}=\int{\frac{\tau_{\nu}d(\nu)}{N}}
\end{equation}
\begin{equation}
{A_{\rm new}}={ A_{\rm known}\frac{\tau_{\rm new}}{\tau_{\rm known}}}
\end{equation}
To determine $A$ in a matrix of different molecules we used the umbrella mode of NH$_{3}$ at 1070 cm$^{-1}$ ($A$$_{\rm NH_{3}} = 1.7\times10^{-17}$ cm molecule$^{-1}$, d'Hendecourt \& Allamandola 1986) at 15 K as the known reference and assumed the thermally induced acid-base reaction of HNCO and NH$_{3}$ to OCN$^{-}$ and NH$_{4}^{+}$, earlier described by Demyk et al. (1998), to be 100\% efficient. Using exactly the same deposition parameters to compose the ice mixtures made it possible to correct for a change in the reference band strength of NH$_{3}$ due to the ice matrix composition. The obtained band strengths are listed in Table~\ref{bs}. Within the accuracy of our measurement, 20\%,  band strengths are in good agreement with those derived earlier (Gibb et al. 2000; d'Hendecourt et al. 1986; Tegler et al. 1995; Lowenthal et al. 2002; Schutte \& Khanna 2003).
\begin{table}
\caption{Band strength of relevant features at 15 K}
\label{bs}
\begin{tabular}{lll}
\hline
& $\nu$ (cm$^{-1}$) & A (molec.cm$^{-1}$)$^{a}$ \\
\hline
\multicolumn{3}{l}{H$_{2}$O dominated matrix: H$_{2}$O/NH$_{3}$/HNCO = 120/10/1}\\\\
HNCO & 2260 & 7.2$\times10^{-17}$  \\
OCN$^{-}$ & 2170 & 1.3$\times10^{-16}$ \\
NH$_{4}^{+}$ & 1485 & 4.1$\times10^{-17}$ \\
\hline
\multicolumn{3}{l}{H$_{2}$O poor matrix: NH$_{3}$/HNCO = 10/1}\\\\
HNCO & 2260 & 7.8$\times10^{-17}$ \\
OCN$^{-}$ & 2160 & 1.3$\times10^{-16}$ \\
NH$_{4}^{+}$ & 1485 & 4.6$\times10^{-17}$ \\
\hline
\multicolumn{3}{l}{\footnotesize{a. Band strengths are determined with 20\% accuracy}}\\
\end{tabular}
\end{table}
\subsection{Thermal formation of OCN$^{-}$}
\begin{figure}
\centering
\includegraphics[width=9cm]{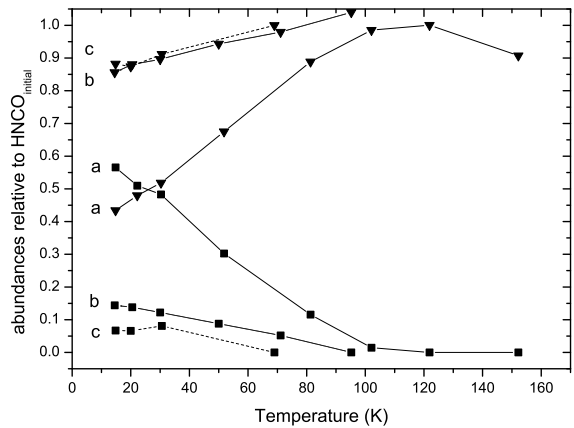}
\caption{The conversion of HNCO to OCN$^{-}$ as a function of temperature is shown quantitatively for $a.$ H$_{2}$O/NH$_{3}$/HNCO = 120/10/1, $b.$ H$_{2}$O/NH$_{3}$/HNCO = 24/10/1 and $c.$ NH$_{3}$/HNCO = 10/1. Abundances are normalised to the initial HNCO abundance. OCN$^{-}$ is marked by triangles while squares indicate HNCO. Conversions can exceed 100\% due to an increased signal intensity of OCN$^-$ at $>$70 K and a 20\% accuracy of the measurements.}
\label{thermalevol2}
\end{figure}

The acid-base like reaction mechanism in which HNCO donates a proton to NH$_{3}$ is the last step in the reaction towards OCN$^{-}$ formation. As this is likely to be effected by the mobility of the reacting species, OCN$^{-}$ formation was studied qualitatively and quantitatively by thermal processing. Three different ice compositions were used to simulate interstellar abundances. The first sample has relative abundances such that they reflect column densities as observed towards W\,33\,A (Gibb et al. 2001). This ice is highly dominated by H$_{2}$O (typical composition H$_{2}$O/NH$_{3}$/HNCO = 120/10/1). Second, a matrix was composed that lacked H$_{2}$O, creating the most favourable extreme environment to form OCN$^{-}$ from HNCO and NH$_{3}$. Third, an in-between sample was taken with a reduced amount of H$_{2}$O, simulating a situation of high local HNCO and NH$_{3}$ abundances.\\ 
%fig1
\begin{figure}
\centering
\includegraphics[width=9cm]{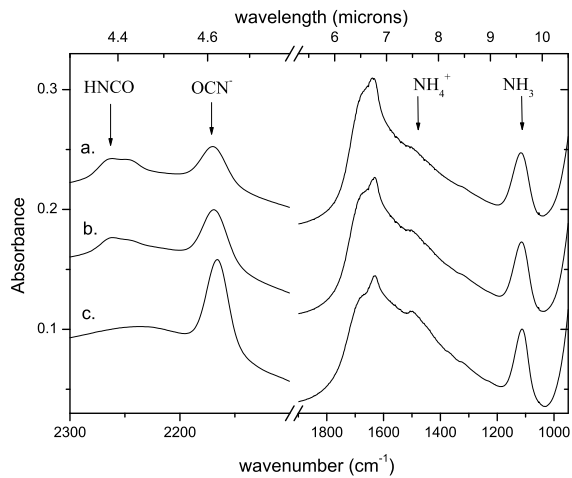}
\caption{The thermal evolution of H$_{2}$O/NH$_{3}$/HNCO = 120/10/1 ice is shown qualitatively at $a.$ $T$ = 15 K, $b.$ $T$ = 52 K and $c.$ $T$ = 122 K in the spectral range of 2300--2100 cm$^{-1}$ and 1850--990 cm$^{-1}$. At 122 K OCN$^{-}$ has reached maximum production.}
\label{thermalevol1}
\end{figure}
\indent The thermal evolution of this acid-base reaction is shown qualitatively in Fig.~\ref{thermalevol1}, which contains spectral regions with the most distinctive features. The strong H$_{2}$O stretch vibration around 3300 cm$^{-1}$ is left out for practical reasons. The unresolved double-peaked structure at 2265 cm$^{-1}$, extending to 2245 cm$^{-1}$, is assigned to HNCO (Novozamsky et al. 2001). As the temperature increases this structure evolves to a single peak at 2260 cm$^{-1}$ which is consistent with the thermal behaviour of HNCO in a H$_{2}$O matrix (Raunier et al. 2003a). At 15 K OCN$^{-}$ is already observed by its feature at 2169 cm$^{-1}$, supported by the counter-ion NH$_{4}^{+}$ that is seen as a shoulder at 1485 cm$^{-1}$ on top of the broad H$_{2}$O bending mode. This is in agreement with findings from Raunier et al. (2003b) who show that OCN$^-$ forms from the interaction of HNCO with 4 NH$_3$ molecules by purely solvation-induced dissociative ionization at 10 K. The features of OCN$^{-}$ and NH$_{4}^{+}$ continue to grow with temperature as the broad HNCO-structure at 2260 cm$^{-1}$ and the NH$_{3}$ feature at 1070 cm$^{-1}$ decrease, consistent with previous experiments by Demyk et al. (1998). Furthermore, as the temperature rises the peak position of OCN$^{-}$ shifts from 2170.8 cm$^{-1}$ at 15 K to 2166.0 cm$^{-1}$ at 120 K.\\
\indent A quantitative analysis is shown in Fig.~\ref{thermalevol2} for the three ices under study. The initial conversion at 15 K is influenced by the relative concentrations of HNCO and NH$_{3}$ and ranges from 40\% in a H$_{2}$O-dominated matrix to a maximum of 90\% in the absence of H$_{2}$O. A conversion at such low temperatures is likely to be partly facilitated by the kinetic energy freed when the molecules freeze out onto the surface. However, studies by Raunier et al. (2003a, b) on the thermal reactivity of HNCO with H$_2$O and NH$_3$ individually show that dissociative ionization of HNCO in H$_2$O only occurs above 130 K but already at 10 K in the presence of NH$_3$. The direct continuation of OCN$^-$ formation that is observed when the temperature is only slightly raised, can be explained by an increasing mobility of NH$_3$ to facilitate the deprotonation of HNCO within the time of our experiment ($\sim$10 min.). Nevertheless, the temperature at which maximal conversion is achieved depends only slightly on the H$_{2}$O dilution. All matrices show the expected conversion of $\sim$100\% for HNCO between 70--100 K that is irreversible upon cooling.\\
\indent The thermal deprotonation of HNCO by H$_2$O, shown by Raunier et al. (2003a), is not observed while UV-induced deprotonation is (see Sect. 3.3). The IR-analysis technique and the lower resolution used in the present study might account for this discepancy. Our observations subscribe that NH$_3$ is not a priori required for this reaction. However, due to its efficiency the reaction of HNCO with NH$_3$, shown in Table~\ref{reactions}, is most likely responsible for the thermal formation of OCN$^-$.\\
%fig2
\begin{table}
\caption{Reactions responsible for OCN$^-$ formation}
\label{reactions}
\begin{tabular}{l}
\hline
{\it{reactions in thermally processed ices}}\\\\
HNCO + NH$_{3}$ $\longrightarrow$ OCN$^{-}$ + NH$_{4}^{+}$ \\ 
HNCO + H$_{2}$O $\longrightarrow$ OCN$^{-}$ + H$_{3}$O$^{+}$ (?)\\\\
{\it{reactions in photoprocessed ices}}\\\\
NH$_{3}$ + $h\nu$ $\longrightarrow$ NH$_{2}$ + H      \\ 
H$_{2}$O + $h\nu$ $\longrightarrow$ OH + H \\
CO + NH$_{2}$ $\longrightarrow$ CONH$_{2}^{*}$ $\longrightarrow$ HNCO + H\\ 
\ \ \ \ \ \ \ \ \ \ \ \ \ \ \ \ $\longrightarrow$ CONH$_{2}$\\
CO + OH $\longrightarrow$ COOH$^{*}$ $\longrightarrow$ CO$_{2}$ + H\\ 
\ \ \ \ \ \ \ \ \ \ \ \ \ \ \ \ $\longrightarrow$ COOH\\
CONH$_{2}$ + H $\longrightarrow$ HCONH$_{2}$ \\
HNCO + NH$_{3}$ $\longrightarrow$ OCN$^{-}$ + NH$_{4}^{+}$ \\
HNCO + H$_{2}$O $\longrightarrow$ OCN$^{-}$ + H$_{3}$O$^{+}$\\
\hline
\end{tabular}
\end{table}
\subsection{UV photoprocessing of HNCO-containing ices}
\begin{figure}
\centering
\includegraphics[width=9cm]{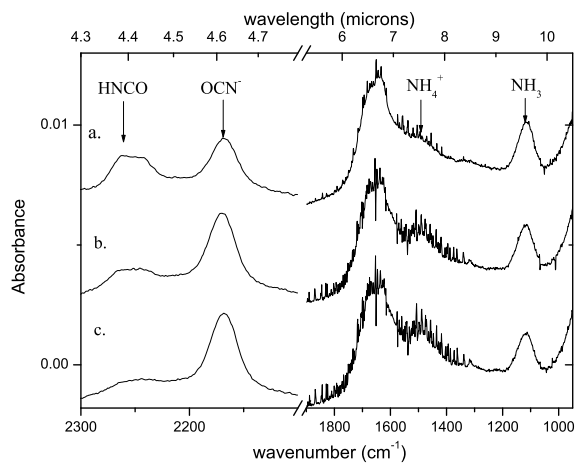}
\caption{The deprotonation of HNCO, induced by `hard' UV photons at 15 K, is shown qualitatively for H$_{2}$O/NH$_{3}$/HNCO = 140/8/1 for the spectral range of 2300--2100 cm$^{-1}$ and 1850--990 cm$^{-1}$ at a UV fluence of $a.$ 0 photons cm$^{-2}$, $b.$ 2.4$\times10^{16}$ photons cm$^{-2}$ and $c.$ 1.0$\times10^{17}$ photons cm$^{-2}$, corresponding to the maximum OCN$^{-}$ produced. CO$_{2}$ appears as the only observable by-product of OCN$^{-}$ formation but is not shown for practical reasons.}
\label{irr_evol2}
\end{figure}

\begin{figure}
\centering
\includegraphics[width=9cm]{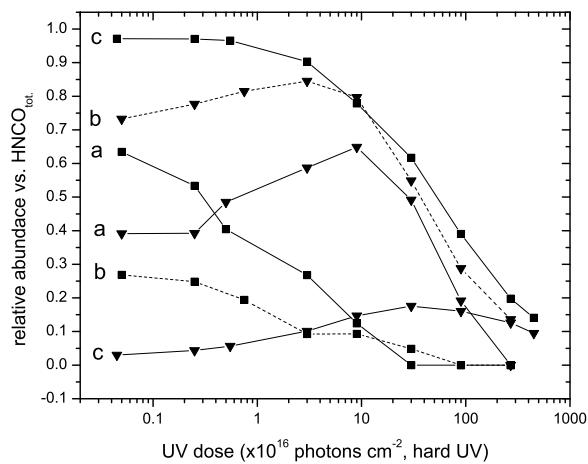}
\caption{The deprotonation of HNCO at 15 K is shown quantitatively as a function of the fluence of `hard' UV photons for $a.$ H$_{2}$O/NH$_{3}$/HNCO = 140/8/1, $b.$ H$_{2}$O/NH$_{3}$/HNCO = 20/8/1 and $c.$  H$_{2}$O/HNCO = 14/1. Abundances are normalised to the initial HNCO abundance. Triangles and squares indicate OCN$^{-}$ and HNCO, respectively. Maximal OCN$^{-}$ abundances are obtained at a fluence of 0.2--3.0$\times10^{17}$ photons cm$^{-2}$.}
\label{irr_evol1}
\end{figure}

The UV photoprocessing of HNCO-containing ices was performed to compare the efficiency of the UV-induced to the thermally induced chemistry discussed above. Similar ice mixtures were grown with compositions H$_{2}$O/NH$_{3}$/HNCO = 140/8/1 (hereafter mixture A), H$_{2}$O/NH$_{3}$/HNCO = 20/8/1, and H$_{2}$O/HNCO = 14/1.\\
\indent The photoprocessing of mixture A at 15 K is shown qualitatively in Fig.~\ref{irr_evol2}. Only the 2300--2100 cm$^{-1}$ and the 1850--950 cm$^{-1}$ regions are shown. Part of the deposited HNCO reacts directly to OCN$^{-}$ (see Sect. 3.2). UV irradiation causes the 2169 cm$^{-1}$ and 1485 cm$^{-1}$ features to increase further in intensity as the 2260 cm$^{-1}$ and 1070 cm$^{-1}$ bands decrease, showing the formation of OCN$^{-}$ and NH$_{4}^{+}$ from HNCO and NH$_{3}$ (described by the reactions in Table ~\ref{reactions}). Apart from CO$_2$ no other photoproducts are seen to form based on the infrared spectroscopy. All nitrogen-bearing species disappear from the solid ice matrix at sufficiently high UV dose. This will be discussed in Sect. 3.4.\\
\indent Fig.~\ref{irr_evol1} shows the reaction of HNCO to OCN$^{-}$ quantitatively as a function of UV dose for the three ices studied. Abundances are given in percentage of the initially deposited amount of HNCO. The maximal efficiency at which OCN$^{-}$ can be produced depends on the H$_{2}$O dilution of the ice matrix. Higher concentrations of NH$_{3}$ and HNCO with respect to H$_{2}$O result in higher efficiencies of OCN$^{-}$ formation. The efficiency ranges from 60\% to 85\% with respect to initial HNCO for the matrices here studied and a maximal OCN$^{-}$ abundance was obtained after a dose of 0.2-3.0$\times10^{17}$ photons cm$^{-2}$. This is $\sim$10$\times$ lower than the amount of photons needed to reach maximal abundances in H$_{2}$O/CO/NH$_{3}$-precursor matrices (see section 3.4). At higher dose, OCN$^{-}$ photodissociation dominates over its formation and after $\sim$2$\times10^{18}$ photons cm$^{-2}$, OCN$^{-}$ can no longer be detected. The thermal and UV processing have very similar effects in H$_{2}$O/NH$_{3}$/HNCO matrices (compare Fig.~\ref{thermalevol1} and ~\ref{irr_evol2}). Photoproducts, formed with a kinetic energy in excesses of the ambient ice temperature, will increase the effective ice temperature when they release this excess energy into the matrix phonon modes. It is therefore not possible to distinguish between these two routes. 
%fig3
%fig4
\subsection{UV photoprocessing of H$_{2}$O/CO/NH$_{3}$ ices}
\begin{figure}
\centering
\includegraphics[angle=90, width=8cm]{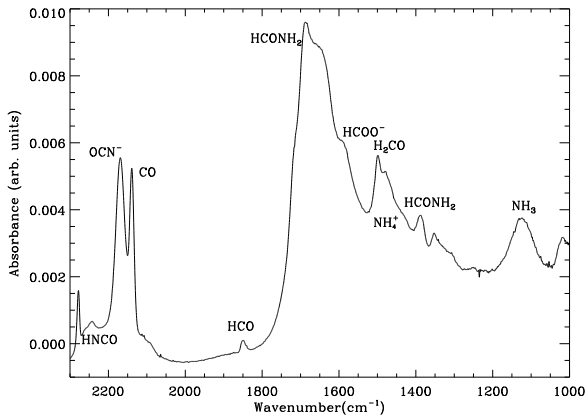}
\caption{The 2300--1000 cm$^{-1}$ spectral range of a standard H$_{2}$O/CO/NH$_{3}$ ice matrix after photoprocessing with `hard' UV photons. The complete spectrum is dominated by the CO$_2$ stretch vibration at 2342 cm$^{-1}$, left out for practical reasons.}
\label{UV}
\end{figure}

\begin{figure}
\centering
\includegraphics[width=9cm]{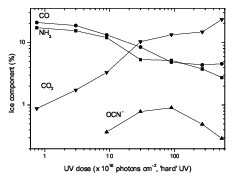}
\caption{Photoprocessing of H$_{2}$O/CO/NH$_{3}$ = 100/22/18 ice with `hard' UV photons. All ice components are shown in percentage of the initial H$_{2}$O abundance.}
\label{hardUV}
\end{figure}

Previous work by Grim \& Greenberg (1987) showed the formation of OCN$^{-}$ by UV photoprocessing of ices composed of H$_{2}$O, CO and NH$_{3}$. The resulting spectrum nicely fits the observed 4.62\,$\mu$m feature toward various lines of sight. A careful interpretation of the interstellar data requires that the interstellar OCN$^{-}$ feature is reproduced not only by shape but also in absolute intensity and that the additional product formation in studied.\\
\indent Apart from OCN$^-$, many products are formed by UV-photon induced reactions in a H$_{2}$O/CO/NH$_{3}$ ice matrix. The 2300--1000 cm$^{-1}$ spectral region in Fig.~\ref{UV} shows the most distinctive features. Most products have been identified in earlier irradiation experiments of similar ices and are listed in Table~\ref{assignment} (Grim et al. 1989; d'Hendecourt et al. 1986; Hagen et al. 1983). The strong stretch vibration of CO$_{2}$ at 2340 cm$^{-1}$ dominates the spectrum but is not shown for practical reasons. A weak, broad, previously unidentified feature appears at 2260 cm$^{-1}$. This feature is partly blended with the 2278 cm$^{-1}$ resonance of CO$_{2}$ and is positively identified with the $\nu{_{2}}$ vibration of HNCO diluted in H$_{2}$O ice. None of the features, listed in Table~\ref{assignment}, form an unique diagnostic for UV photoprocessing in this spectral region because they are either non-specific (CO$_{2}$) or have too weak band strengths to give constraints (H$_2$CO and HCONH$_2$).\\ 
\indent Figure~\ref{hardUV} shows the photolysis of CO and NH$_{3}$ in a H$_{2}$O dominated ice matrix together with the evolution of OCN$^{-}$. Using Equation (1), the derived abundances are normalised to the original H$_{2}$O ice content. For OCN$^{-}$ the absolute integrated band strengths listed in Table ~\ref{bs} are used. CO$_{2}$ is formed almost instantly upon UV exposure. OCN$^{-}$, being a higher order product of irradiation, is detected only after prolonged irradiation and reaches a maximum abundance at a UV dose of $\sim1\times10^{18}$ photons cm$^{-2}$. At higher UV doses it again decreases due to photodestruction. At the highest doses only CO$_{2}$ is seen to form. The nitrogen-bearing species that were detected degrade and most likely form infrared-weak or inactive molecules. Molecular nitrogen or complex large organics are suggested to form in this context (Agarwal et al. 1985; Bernstein et al. 1995; Mu\~noz-Caro \& Schutte 2003). \\
\indent Table ~\ref{peak} shows a summary of the ice matrices used to investigate the efficiency of the OCN$^{-}$ formation. The repeatability of the listed values is typically of the order of 20\%. The formation efficiency rises when the fraction of CO and NH$_3$ in the ice is increased, whereas the elimination of H$_{2}$O from the matrix shows no significant effect. In order to form OCN$^{-}$ at the abundance observed towards W\,33\,A, the minimum initial ice fraction of CO and NH$_{3}$ must be of the order of 30\% or higher. No ice was found to form more than 2.7\% OCN$^{-}$. On average the production efficiency ranges between 0.6\% and 1.8\%. \\
%fig5
\begin{table}
\caption{Peak assignment of a standard H$_{2}$O/CO/NH$_{3}$ ice after UV-photoprocessing.}
\label{assignment}
\begin{tabular}{lll}
\hline
Wavenumber (cm$^{-1}$) & Assignment$^{a}$ & A (cm$^{2}$ molec$^{-1}$)\\
\hline
\multicolumn{3}{l}{\it{Initial ice matrix}}\\
3290 & H$_{2}$O & 2.0$\times10^{-16 d}$\\
2139 & CO & 1.0$\times10^{-17 d}$\\
1127 & NH$_{3}$ & 1.3$\times10^{-17 d}$\\\\
\multicolumn{3}{l}{\it{Photoproducts}}\\
2342 & CO$_{2}$ & 7.6$\times10^{-17 e}$\\
2278 & $^{13}$CO$_{2}$ & \\
2260 & HNCO & 7.2$\times10^{-17 b}$\\
2169 & OCN$^{-}$ & 1.3$\times10^{-16 b}$\\
1849 & HCO$^{c}$ &\\
1720 & H$_{2}$CO & \\
1690 & HCONH$_{2}$ & 3.3$\times10^{-17 i}$\\
1580 & HCOO$^{-}$ & 1.0$\times10^{-16 h}$\\
1499 & H$_{2}$CO$^{2}$ & 3.9$\times10^{-18 f}$\\
1485 & NH$_{4}^{+}$ & 4.1$\times10^{-17 b}$\\
1386 & HCONH$_{2}$ & 2.8$\times10^{-18 h}$\\
1352 & HCOO$^{-}$ & 1.7$\times10^{-17 h}$\\ 
1330 & HCONH$_{2}$ & \\
1020 & CH$_{3}$OH & 1.9$\times10^{-17}$$^{g}$\\
650  & CO$_{2}$ & \\
\hline
\end{tabular}
\footnotesize{$^a$ Assignment from Grim et al. (1989) unless otherwise noted; $^b$ This article; $^c$ d'Hendecourt et al. (1986); $^d$ d'Hendecourt \& Allamandola (1986); $^e$ Gerakines et al. (1995); $^f$ Schutte et al. (1993); $^g$ Schutte et al. (1991); $^h$ Schutte et al. (1999);$^i$ Wexler (1967)}
\end{table}
%fig6
%fig7
\begin{figure}
\centering
\includegraphics[width=9cm]{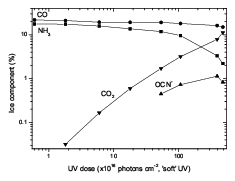}
\caption{Photoprocessing of H$_{2}$O/CO/NH$_{3}$ = 100/22/18 ice with `soft' UV photons. All ice components are shown in percentage of the initial H$_{2}$O abundance.}
\label{softUV}
\end{figure}
%fig8
\begin{figure}
\centering
\includegraphics[width=9cm]{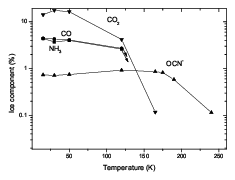}
\caption{Thermal annealing of a photoprocessed H$_{2}$O/CO/NH$_{3}$ ice (`hard' UV).All ice components are shown in percentage of the initial H$_{2}$O abundance.}
\label{heat}
\end{figure}
\indent Standard experimental conditions used a `hard' UV-irradiation source. The influence of the photon characteristics on the formation of OCN$^{-}$ was tested with a less intense, less energetic `soft' UV source that contains primarily photons of $\sim$160 nm. `Soft' processing is shown in Fig.~\ref{softUV}. The photoproducts formed are similar as those due to `hard' UV. However, due to the difference in energy distribution, products appear at different doses. Under `soft' conditions, the maximum OCN$^{-}$ formation occurs at a dose of $\sim4\times10^{18}$ photons cm$^{-2}$, a factor of 4 higher than in the `hard' environment, but its maximum abundance is hardly effected, 1.1\% versus 0.8\%. The destruction of CO and the formation of CO$_{2}$ decrease enormously with `soft' UV but the evolution of NH$_{3}$ is almost unaltered: only 20\% of its initial abundance is left when OCN$^{-}$ reaches maximum abundance. This dependence on UV photon energy is consistent with the wavelength dependence of the photodissociation cross sections of these species, with CO being destroyed only at $\lambda$ $<$ 110 nm (van Dishoeck 1988).\\
\indent When a photoprocessed matrix is thermally annealed, the OCN$^{-}$ production efficiency seems to increase. In Fig.~\ref{heat} the dominant ice components of a photoprocessed H$_{2}$O/CO/NH$_{3}$ ice, plotted relative to the initial H$_{2}$O abundance shows an increase of the 4.62\,$\mu$m signal intensity as a function of temperature to a maximum of $\sim$12\% at 180 K. UV-photolysis experiments have shown that small amounts of HNCO form in the ice, contributing to an increasing signal by thermal conversion to OCN$^{-}$ at temperatures of 70--100 K. However, since the increase is prominent only at higher temperatures while thermal deprotonation is already apparent below 70 K, it is more likely induced by changing interactions in the OCN$^{-}$--NH$_{4}^{+}$ salt-complex. 
NH$_{3}$, CO and CO$_{2}$ evaporate in the laboratory between 120 K and 160 K when mixed in H$_2$O. OCN$^{-}$ remains present in the condensed phase up to $\sim$200--240 K and the small strengthening of the 4.62\,$\mu$m signal intensity is accompanied by the evaporation of the H$_{2}$O ice, motivating the argument of salt formation. \\  
\begin{table}
\caption{OCN$^{-}$ peak abundances obtained by UV photoprocessing}
\begin{center}
\label{peak}
\begin{tabular}{lllll}
\hline
\multicolumn{3}{l}{Matrix composition}&\multicolumn{1}{l}{Layer}& \multicolumn{1}{l}{OCN$^{-}$$^{a}$}\\
& & & thickness & (max.)\\
H$_{2}$O & CO & NH$_{3}$ & ($\mu$m) &  \\
\hline\\
\multicolumn{3}{l}{Variable layer thickness} & &  \\ \\
100 & 24 & 22 & 0.026 & 1.03  \\
100 & 24 & 20 & 0.08 & 1.26 \\
100 & 24 & 21 & 0.12 & 1.38 \\ \\
\multicolumn{3}{l}{Variable NH$_{3}$ abundance} & &  \\ \\
100 & 24 & 9 & 0.08 & 0.50 \\ 
100 & 23 & 31 & 0.03 & 1.74 \\ 
100 & 23 & 34 & 0.14 & 1.98 \\ \\
\multicolumn{3}{l}{Variable CO abundance} & &  \\ \\ 
100 & 8 & 20 & 0.06 & 0.58 \\ 
100 & 39 & 20 & 0.011 & 1.29 \\ 
100 & 55 & 21 & 0.08& 1.78\\
 - & 56 & 20& 0.07& 1.77\\ \\
\multicolumn{3}{l}{Variable NH$_{3}$ at double CO } & &  \\ 
\multicolumn{3}{l}{abundance} & &  \\\\ 
100 & 56 & 8 & 0.08 & 0.45 \\
100 & 53 & 32 & 0.08 & 2.68 \\ 
100 & 42 & 32 & 0.05 & 1.88 \\\\ 
\multicolumn{3}{l}{Variable NH$_{3}$ at half CO} & &  \\ 
\multicolumn{3}{l}{abundance} & &  \\\\ 
100 & 9 & 8 & 0.07 & 0.32 \\ 
100 & 7 & 31 & 0.07 & 0.56 \\\\ 
\multicolumn{3}{l}{H$_{2}$O/CH$_{3}$OH/NH$_{3}$} & &  \\ \\
100 & 22 & 9 & 0.09 & 1.1  \\
100 & 35 & 11 & 0.07 & 1.6  \\
100 & 15 & 15 & 0.10 & 1.2  \\
100 & 28 & 40 & 0.07 & 1.6  \\\\
\multicolumn{3}{l}{Various} & &   \\ \\ 
100 & 24 & 20 $^{b}$ & 0.07 & 0.96 \\ 
100 & 25 & 20 $^{c}$& 0.10 & 1.05 \\ 
100 & 50 & 20 $^{d}$& 0.016 & 0.41\\
100 & - & 18 $^{e}$& 0.12 & 1.00 \\ 
- & 24 & - $^{f}$& 0.08& -\\
- &- & 20 $^{g}$& 0.05& -\\
- &- & - $^{h}$& 0.06& -\\ 
\hline
\end{tabular}
\end{center}
\footnotesize{All experiments are performed under standard conditions, $T$ = 15 K and irradiation with `hard' UV photons of 4$\times10^{4}$ K, unless otherwise stated. Results are obtained with an accuracy of 20\%.; $^a$ abundance relative to the originally abundant H$_{2}$O in \%; $^b$ $T$ = 50 K; $^c$ Irradiation with a UV field of soft UV photons (1$\times10^{4}$ K) at $T$ = 15 K; $^d$ H$_{2}$O/NH$_{3}$ layer (0.01 $\mu$m) with a CO layer (0.006 $\mu$m) on top; $^e$ H$_{2}$CO (28\%) substitutes CO; $^f$ N$_{2}$ (24\%) substitutes NH$_{3}$; $^g$ CH$_{4}$ (20\%) substitutes CO; $^h$ N$_{2}$ (24\%) and CH$_{4}$ (20\%) substitute NH$_{3}$ and CO.}
\end{table}
\indent All the matrices studied were optically thin to UV radiation. Nevertheless, thicker ice layers are seen to support a higher OCN$^{-}$ formation efficiency than thinner ones. The thicker the ice, the lower the ratio of photons/molecule needed to favor the processes involved in formation. A second, more plausible, explanation could be a caging effect that effectively slows down the dissociation of HNCO into NH + CO, making the reaction of H + NCO more prominent. This is shown in noble gas matrices by Pettersson et al. (1999). Since the photoproduction of OCN$^{-}$ is also likely to be matrix-assisted and the amorphous ice in which OCN$^{-}$ is formed allows for some mobility of the photoproducts, the trapping of volatile intermediate products will be more effective in thicker ice layers and will result in a higher maximal OCN$^{-}$ abundance. Our experiments are, however, not detailed enough to draw stronger conclusions on the mechanisms involved.  \\
\indent Under {\it{'Various'}} in Table~\ref{peak} some experiments are listed that differ from the adopted standard conditions. Either the experimental conditions or the initial composition of the matrix is changed. UV processing of ice at elevated temperatures of 50 K results in a lower OCN$^{-}$ production efficiency than at 15 K. The ice was grown under standard conditions at 15 K and then brought to 50 K for irradiation. Fig.~\ref{heat} shows that CO and NH$_{3}$ slowly evaporate from such an ice matrix, which will cause a continuous decrease in concentration that in turn leads to a lower photoproduction of OCN$^{-}$. Alternatively, volatile intermediates evaporate more easily at higher temperatures, leading to lower production. This observation supports the argument above that increased product mobility in the ice lowers the maximum OCN$^{-}$ formation efficiency.\\
\indent A two phase model, in which CO is deposited on top of H$_{2}$O and NH$_{3}$, shows that NH$_{3}$ and CO residing in different, adjoining, ice layers on a grain forms no restriction for OCN$^{-}$ formation. The efficiency of formation is considerably lower than for mixed matrices but much higher than would be expected from a purely interface-based reaction and could be partly explained by CO grown on top of amorphous ice, making use of a much larger surface area to react. Also, UV photons are likely to locally increase the effective temperature of the ice that increases diffusion and can lead to the mixing and trapping of CO, such as occurs upon thermal warming.\\
\indent In analogy to CO, the more reduced species H$_{2}$CO, CH$_{3}$OH and CH$_{4}$ were also tested for their OCN$^{-}$ production efficiency. H$_{2}$CO and CH$_{3}$OH showed favorable results and will be discussed in Sect. 3.5. CH$_{4}$ was found not to be an OCN$^{-}$ precursor through UV photoprocessing. Also when NH$_{3}$ was substituted by N$_{2}$ no OCN$^{-}$ formation was observed.
%fig9
\begin{figure}
\centering
\includegraphics[width=9cm]{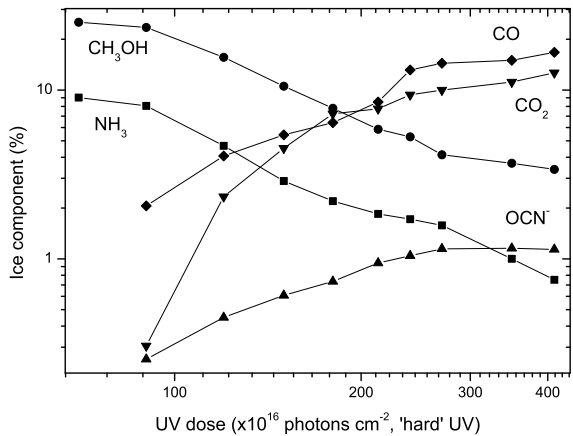}
\caption{Photoprocessing of H$_{2}$O/CH$_{3}$OH/NH$_{3}$ = 100/22/9 ice with `hard' UV photons. All ice components are shown in percentage of the initial H$_{2}$O abundance.}
\label{meth}
\end{figure}
\subsection{UV photoprocessing of H$_{2}$O/CH$_{3}$OH/NH$_{3}$ ice}
In addition to the CO photolysis experiments, the OCN$^{-}$ production has been examined in several ice matrices in which CO was replaced by more reduced analogs, e.g. H$_{2}$CO and CH$_{3}$OH. Methanol is the second most abundant molecule observed in interstellar ices for some lines of sight (Allamandola et al. 1992; Dartois et al. 1999a; Pontoppidan et al. 2003). Previous experiments by Bernstein et al. (1995) showed that photoprocessing of methanol in H$_{2}$O-ice produces similar oxygen-containing molecules as ices containing H$_{2}$O, CO and NH$_{3}$. A further look into the production efficiency from CH$_{3}$OH at various relative abundances shows that less NH$_{3}$ is required in the starting ice-matrix to produce similar amounts of OCN$^{-}$ as does a corresponding CO-containing ice matrix. Fig.~\ref{meth} shows the photoprocessing of NH$_{3}$ and CH$_{3}$OH and the evolution of OCN$^{-}$. Compared to Fig.~\ref{hardUV} both ices form similar quantities of OCN$^{-}$, however, in the presence of CH$_{3}$OH only half as much NH$_{3}$ is required initially to obtain the same amount of OCN$^{-}$ (also Table ~\ref{peak}).  Notably, in this case, a higher ($\sim$3$\times$) fluence of $\sim$4$\times10^{18}$ `hard' UV-photons cm$^{-2}$ is needed to reach maximum OCN$^{-}$ abundance, resulting in higher NH$_{3}$ destruction. \\
%fig10
\begin{figure}
\centering
\includegraphics[width=9cm]{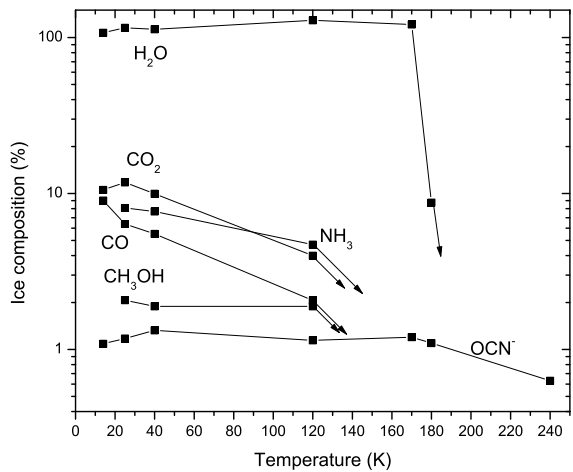}
\caption{Thermal annealing of H$_{2}$O/CH$_{3}$OH/NH$_{3}$ = 100/15/15 ice. All species are in percentage relative to the initial H$_{2}$O abundance.}
\label{45b}
\end{figure}
\indent The thermal processing of a photoprocessed H$_{2}$O/CH$_{3}$OH/NH$_{3}$ = 100/15/15 ice (initial composition) at maximal OCN$^{-}$ production  is shown in Fig.~\ref{45b} and looks highly similar to the thermal processing of its irradiated CO-analog shown in Fig.~\ref{heat}. H$_{2}$O is included in this plot as a percentage of its own initial band strength to give an idea of the coexistence of H$_{2}$O-ice and solid OCN$^{-}$ at different temperatures. It is important to realize that the H$_{2}$O-ice feature at 3\,$\mu$m undergoes an enormous change around 120 K due to crystallisation and that this will directly effect the OCN$^-$ matrix environment. It is also interesting to note that OCN$^{-}$ evaporates at somewhat higher temperatures, 200--240 K, than H$_{2}$O but that the evaporation of H$_{2}$O is not accompanied by a major change in the infrared-feature at 4.62\,$\mu$m.      
\begin{table*}[htb]
\caption{Abundances and temperatures observed toward various sources}
\begin{center}
\label{source1}
\begin{tabular}{lccccccc}   \hline \hline
Source & A$_v$ & H$_2$O & NH$_3$/H$_2$O &  CH$_3$OH/H$_2$O & OCN$^-$/H$_2$O & CO$_{total}$& CO$_2$-bend \\ 
  &  & (10$^{18}$) & (10$^{-2}$) &   (10$^{-2}$) & (10$^{-3}$)&$N\times10^{17}$cm$^{-2}$ & T (K) \\  \hline \hline  
W3 IRS5 & 144$^{16}$ & 5.4$^1$ & $<$3.5$^1$ &   $\leq 2.96$$^{16}$ & $\leq 5.2$$^{34}$ &$<$1.7$^{17}$& 136$^1$\\
W\,33\,A & 148$^{14}$ & 12$^{15}$ & $\leq 5$$^{31}$   & 15.4$^{14}$ & 22.1$^{15}$ &3.9$^{17}$& 136$^1$/50--90$^{19}$ \\
GL 2136 & 94.3$^{37}$  & 5$^1$ &5.4$^{25}$    & 5.2$^{20}$ & 2.6$^{27}$ &0.62$^{26}$& 117$^1$ \\
GL 2591 & n.d.  & 1.7$^1$   &n.d. & 4.12$^{1}$ &n.d. &$<$0.4$^{26}$& 117$^1$ \\ 
S140 & 75$^{16}$ & 2.15$^1$ & n.d. &  $\leq 2.79$$^{16}$ &n.d. &n.d.& 136$^1$ \\
NGC 7538 IRS9 & $\geq 84^{16}$ & 8$^1$ & 8$^{25}$ &   6.06$^{23}$ & 5.8$^{27}$ &9.6$^{26}$& 119$^1$/50--90$^{19}$ \\
AFGL 7009S & 75$^{14}$ & 12$^{14}$ & -$^{11}$    & 29.5$^{14}$ & 13.5 $^{28}$&n.d.& 85$^3$ \\ 
AFGL 961E & 40$^{14}$ & 4.2$^{18}$ & n.d.   & $\leq 4.76$$^{16}$ & 2.7$^{27}$ &n.d.& n.d.\\
MonR2 IRS2 & 21$^{17}$ & 5.9$^{2}$ &  n.d.  & 2.20$^{20}$ & 6.5$^{27}$ &n.d.&n.d.  \\ 
IRAS 08448-4343 & 40$^5$ & 7.8$^5$ &  n.d.  & 0.84$^5$ & 1.7$^{5}$ &n.d.& n.d.\\
\hline 
L1551 IRS5 & $\geq 19^{10}$ & 3.5$^{29}$   & n.d. &n.d. & 14.1$^{29}$ &3.0$^{17}$& n.d. \\
RNO 91 & $\geq 10^7$ & 2.2$^2$ &n.d.   &n.d. & $\leq 13.1$$^{2}$ &n.d.&n.d. \\  
Elias 18 & 17$^6$ & 1.5$^2$ &   n.d. & 5.33 $^{21}$& 7.2$^{30}$ &2.3$^{12}$& 20--30$^{12}$\\
PV Cep & 19$^{32}$ & 7.4$^2$ &n.d. & n.d.  & $\leq 0.8$$^{2}$ &n.d.&n.d. \\
Elias 1-12 & 10$^6$ & 1.1$^2$ & n.d.& n.d.  & $\leq 5.6$$^{2}$ &n.d.& n.d.\\
HL Tau & 22$^{22}$ & 1.4$^2$ & n.d. & n.d.& $\leq 5.1$ $^{21}$& $\leq 2.1$$^{2}$&n.d.\\
HH100-IR & 25$^9$ & 3.9$^{14}$ & 5.6$^{25}$ &   $\leq$ 6.15$^{24}$ & 3.4 $^{4}$&n.d.& 10$^{12}$ \\
Elias 29 & $<$23$^{19}$ & $<$1.1$^{19}$ & $<$13$^{11}$ &   $\leq 2.5$$^{19}$ & $\leq 0.4$$^{19}$ &1.7$^{19}$& $<$40$^{11}$ \\ \hline
Elias 16 & 21$^{32}$ & 2.5$^{12}$ & n.d. &   $\leq 2.9$$^{21}$ & $\leq 23.1$$^{4}$ &6.$^{19}$5& 10$^1$ \\
GC:IRS19 &n.d. & 3.6$^8$ & 5$^{35}$ & n.d.  & 32.1$^{8}$ &1.4$^{13}$& 10--40$^{8}$(H$_{2}$O 3.0\,$\mu$m)\\ 
SgrA* & $\sim$31$^{36}$ & 1.24$^{35}$ & $\leq$5$^1$   & $<$4$^{35}$ & 20$^{33}$ &$<$1.5$^{35}$& 10--15$^{35}$(H$_{2}$O 3.0\,$\mu$m)\\
NGC 4945 & n.d. & 3.5-4.3$^{33}$ &n.d. &n.d.  & 38--46$^{33}$  &9.7$^{33}$&$\leq$90$^{33}$ \\
\hline
\end{tabular} 
\end{center}
\begin{list}{}{}
\item{\footnotesize{n.d. indicates measures that have not yet been determined}; $^1$ Gerakines et al. (1999);
$^2$Keane et al.(2001a);
$^3$ Dartois et al. (1999a);
$^4$ Whittet et al. (2001);
$^5$ Thi et al. (2002);
$^6$ Whittet et al. (1985);
$^7$ Weintroub et al. (1991);
$^8$ Chiar et al. (2002);
$^9$ Whittet et al. (1996);
$^{10}$ Cohen et al. (1975), Davidson \& Jaffe (1984);
$^{11}$ Boogert et al. (2000a);
$^{12}$ Nummelin et al. (2001), Temperature determination by fitting of the 4.27$\mu$m feature of CO$_2$;
$^{13}$ Moneti et al.(2001), Temperature is determined towards Sgr A* from CO gas lines;
$^{14}$ Dartois et al. (1999b);
$^{15}$ Gibb et al. (2000);
$^{16}$ Brooke et al. (1996);
$^{17}$ Tielens et al. (1991)
$^{18}$ Smith et al. (1989);
$^{19}$ Boogert et al. (2000b);
$^{20}$ Brooke et al. (1999);
$^{21}$ Chiar et al. (1996);
$^{22}$ Stapelfelt et al. (1995);
$^{23}$ Allamandola et al. (1992);
$^{24}$ Graham et al. (1998);
$^{25}$ Gurtler et al. (2002);
$^{26}$ Sandford et al. (1988);
$^{27}$ Pendleton et al. (1999);
$^{28}$ Demyk et al. (1998);
$^{29}$ Tegler et al. (1993);
$^{30}$ Tegler et al. (1995);
$^{31}$ Taban et al. (2003);
$^{32}$ Whittet et al. (1988);
$^{33}$ Spoon et al. (2003);
$^{34}$ Lacy et al. (1979);
$^{35}$ Chiar et al. (2000);
$^{36}$ Rieke et al. (1989) and references therein;
$^{37}$ Willner et al. (1982)
}
\end{list}
\end{table*}
%fig11
\begin{figure}[h]
\centering
\includegraphics[width=9cm]{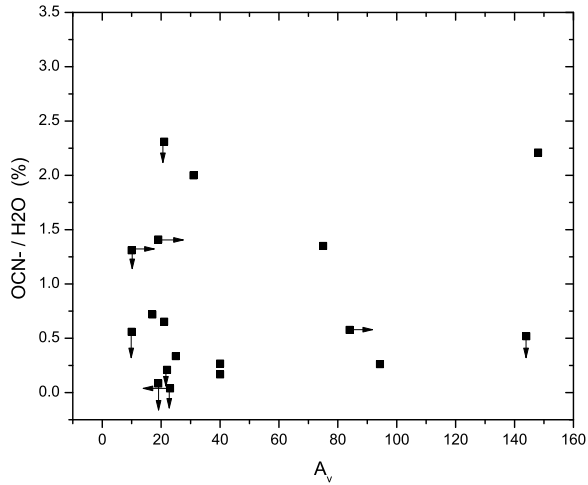}
\caption{Correlation diagram of OCN$^{-}$, shown in abundance relative to H$_{2}$O, with the visual extinction $A_{v}$. Upper limits are indicated by arrows.}
\label{ocn_av}
\end{figure}
\section{Astrophysical implications}
OCN$^{-}$ is easily formed in the laboratory by energetic processing of interstellar ice analogs. Moreover, the fact that OCN$^{-}$ can be readily produced by UV photolysis has led to the direct association of this interstellar feature with UV processing of ices in molecular clouds (Pendleton et al. 1999). The unambiguous interpretation of this feature as a diagnostic of photoprocessing, however, needs some in-depth analysis of the conditions under which OCN$^{-}$ forms, its yield and the by-products of its formation. Hereby the caveat of Sect. 2 must be kept in mind when scaling laboratory data to the observations. \\
\indent Detections and upper limits on the interstellar 4.62\,$\mu$m feature observed towards a variety of lines-of-sight are summarised in Table~\ref{source1}. The top part of the table lists high-mass YSO's, the middle part low-mass YSO's and the bottom part background sources, GC sources and an external galaxy. Typical abundances with respect to H$_2$O range from 0.2\%--0.8\% for both high and low mass stars, with a few exceptions where the observed values increase up to 1.4\%--2.2\%. W\,33\,A, by far the best studied object, shows the highest OCN$^{-}$ content (2.2\% relative to H$_{2}$O) in our galaxy, making it best suited to constrain the possible OCN$^{-}$ production mechanisms. Recently Chiar et al. (2002) detected an even higher abundance of 3.6\% OCN$^{-}$ toward GC:IRS 19, and Spoon et al. (2003) find 3.4\%--3.9\% OCN$^{-}$ towards the central region of the starburst galaxy NGC 4945.  Fig.~\ref{ocn_av} shows that there is no correlation with $A_{v}$, stressing that OCN$^{-}$ is most likely inhomogeneously distributed over the line-of-sight.\\ 
\indent The experiments presented in Sect. 3 clearly show that OCN$^{-}$ can form from many different ices under various conditions. Its formation efficiency is found to depend on the energy distribution of the radiation source and on the thickness, the temperature and, above all, the composition of the ice. Here we assess the probability of the various formation routes under interstellar conditions as well as search for correlations with established interstellar parameters.
\subsection{UV photoprocessing: CO}
UV photoprocessed ice matrices, composed of H$_{2}$O, CO and NH$_{3}$, are shown to match the interstellar 4.62\,$\mu$m feature nicely but a quantitative analysis proves that this route is inefficient. A large amount of CO and NH$_{3}$ is needed to reproduce the observed abundances. This problem is most severe towards W\,33\,A. Assuming that the 2.2\% of OCN$^{-}$ observed towards W\,33\,A corresponds to a local maximum, Table~\ref{peak} shows that the initial ice matrix needs to have at least 25\%--50\% CO and 30\% NH$_{3}$ at 15 K, relative to H$_{2}$O. Table~\ref{source1} includes the observed solid CO and NH$_3$ abundances towards a variety of sources. CO ice is detected towards numerous lines-of-sight and sets no constraints on the production possibility of OCN$^{-}$. NH$_{3}$ is more controversial with upper limits of the order of 5\% (Taban et al. 2003). Even when photodestruction and storage in NH$_4^+$ is taken into account this upper limit makes it unlikely that interstellar ices could contain abundances as high as 30\%. On the other hand, Table~\ref{peak} shows that typical abundances of 0.2\%--0.8\%, found for OCN$^-$ towards a number of sources (Table 5), are produced from a realistic interstellar ice analog, initially containing $\sim$8\% NH$_{3}$ with respect to H$_{2}$O.\\
\indent In addition to NH$_{3}$, the presence of UV photons enforces a constraint on the photoproduction of OCN$^-$. In the laboratory, the UV fluence that produces the maximum amount of OCN$^{-}$ corresponds to a molecular cloud at an age of $\sim$4$\times10^{7}$ yr with a cosmic-ray induced UV-field of 1.4$\times 10^4$ photons cm$^{-2}$ s$^{-1}$ (Prasad \& Tarafdar 1983). This timescale is quite long compared with that estimated for the dense pre-stellar and YSO phases where ices are produced. Regions which could have sufficient UV to form OCN$^{-}$ include the outer regions of protoplanetary discs (Herbig and Goodrich 1986) or clouds located close to bright O \& B stars (Mathis et al. 1983). However, since the 4.62\,$\mu$m absorption band is only observed in association with H$_{2}$O ice, which is generally not seen in regions with large UV intensity, this casts doubt on its formation in such regions as well (Whittet et al. 2001). Based on these arguments, UV photoprocessing of H$_{2}$O, CO and NH$_{3}$ containing ices alone cannot account for the OCN$^{-}$ abundance observed towards W\,33\,A, GC:IRS19, SgrA* and NGC 4945. 
%fig12
\begin{figure}      
\centering
\includegraphics[width=9cm]{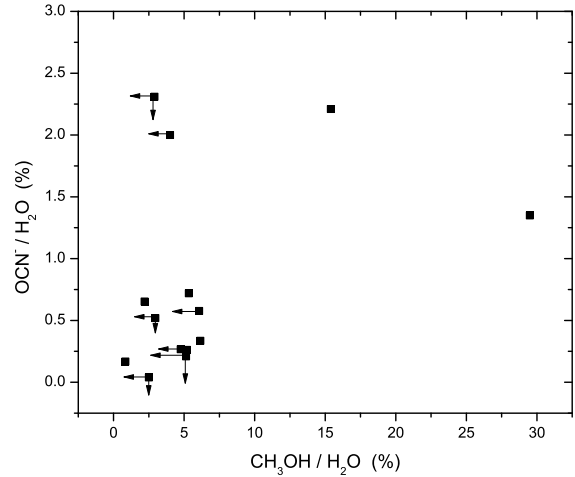}
\caption{Correlation diagram of OCN$^{-}$ and CH$_{3}$OH, both shown in abundance relative to H$_{2}$O. Upper limits are indicated by arrows.}
\label{ocn_ch3oh}
\end{figure}
\subsection{UV photolysis: Methanol}
Methanol is the second most abundant ice component towards W\,33\,A and AFGL 7009S (Dartois et al. 1999a) and is highly abundant in the line-of-sight towards many objects that also show OCN$^{-}$. Gibb et al. (2000) suggest a possible correlation between the extent of ice processing and the methanol abundance. CH$_{3}$OH is thought to form by the sequential hydrogenation of CO on grain surfaces (Tielens \& Charnley 1997), which is believed to include some endothermic reaction steps that could be facilitated, or exclusively occur, when the ice is energetically processed, either thermally or photochemically. If the origin of CH$_{3}$OH is somehow coupled to irradiation, then a correlation with OCN$^{-}$ could be expected. Fig.~\ref{ocn_ch3oh} shows the abundances of CH$_{3}$OH and OCN$^{-}$, relative to H$_{2}$O. A tentative but weak trend can be observed to support this idea, but it is based on only a few clear detections and decidedly more upper limits. This figure does not yet include the new detections of CH$_{3}$OH in low-mass YSOs by Pontoppidan et al. (2003). It will be particularly interesting to check the OCN$^-$ abundance in these objects.\\
\indent Laboratory experiments on the photoprocessing of H$_{2}$O/CH$_{3}$OH/NH$_{3}$ ices, Table~\ref{peak}, show methanol to be a similarly effective precursor of OCN$^{-}$ as CO. The high methanol abundances observed support this possible route of formation. Again a considerable initial abundance of NH$_{3}$ ice is required, although less than in CO mixtures. The UV fluence needed to produce OCN$^{-}$ via CH$_{3}$OH, which is $\sim$3$\times$ higher than for CO, results in a high degee of photodissociation of NH$_{3}$, such that any remaining NH$_{3}$ could be consistent with the observed conservative upper limits (5\% relative to H$_2$O towards W\,33\,A). However, such a high fluence is not expected to be present in the environments that show the 4.62\,$\mu$m feature which puts strong doubts on this formation mechanism (see discussion above).
\subsection{UV photolysis or thermal processing: HNCO}
\begin{figure}[h]
\centering
\includegraphics[width=9cm]{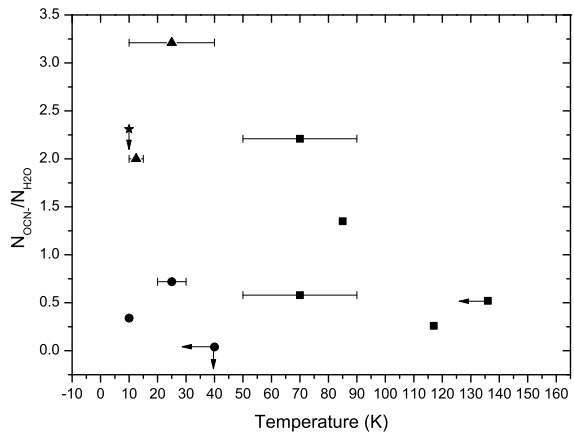}
\caption{OCN$^{-}$ abundance relative to H$_{2}$O, as a function of the ice temperature, observed towards YSOs. Upper limits are indicated by arrows.}
\label{temp}
\end{figure}

The acid-base reaction that forms OCN$^{-}$ in the condensed phase is efficiently mediated by thermal processing as well as by UV photoprocessing of HNCO-containing ice in the laboratory. How HNCO ices are formed on interstellar grains remains an unaddressed issue in this paper. HNCO is abundant in the gas phase in hot cores and regions associated with high mass star formation and shocked gas (Zinchenko et al. 2000 and references therein) but no HNCO has yet been detected in interstellar ices. One way to form HNCO is by grain-surface reactions in dense molecular clouds, which can lead to abundances of $\sim$3\% with respect to H$_2$O (Hasegawa \& Herbst 1993). In the gas phase, either an ion-molecule reaction mechanism or a neutral-neutral reaction pathway is suggested (Iglesias 1977; Turner et al. 1999). \\
\indent We determined upper limits on the HNCO ice abundance for NGC 7538 IRS9 ($\leq$ 0.5\%), GL 2136 ($\leq$ 0.7\%) and W\,33\,A ($\leq$ 0.5\%). These allow for a minimal conversion to OCN$^{-}$ of $\sim$50\%, $\sim$50\% and $\sim$75\%, respectively. NH$_{4}^+$, detected toward various sources, is sufficiently abundant to balance the observed OCN$^{-}$ (Schutte \& Khanna 2003; Keane et al. 2001a). Induced by UV, this reaction mechanism would require a minimal fluence of $\sim2\times10^{16}$ photons cm$^{-2}$ (50\% conversion) or $1\times10^{17}$ photons cm$^{-2}$ (75\% conversion) of hard UV in the laboratory. In analogy to the discussion above on the presence of UV this very likely exceeds the acceptable fluence seen by a grain inside a cloud intergrated over the lifetime of a molecular cloud.\\
\indent Alternatively, a $\sim$75\% conversion can be thermally mediated at a temperature of 60--90 K in the laboratory, which is in good agreement with the temperatures found for the dominant part of the ice observed towards NGC 7538 IRS9, GL 2136 and W\,33\,A. Table 5 includes the (laboratory) temperature that best fits the solid CO$_2$ feature (Boogert et al. 2000b; Gerakines et al. 1999). Figure~\ref{temp} shows the OCN$^{-}$ abundance as a function of this temperature. No OCN$^{-}$ is detected toward sources with CO$_2$ temperatures higher than 140 K but it seems randomly distributed at lower temperatures. Since its thermal formation starts at 15--20 K, this indirectly indicates an initially random HNCO distribution. However, such a conclusion is premature since the presence or absence of other sources of energetic processing will highly influence any relation.\\\indent The combination of the efficiency of this process ($\sim$100\%), the low NH$_3$ requirement, the low temperature formation route and the detection of NH$_4^+$ towards W\,33\,A make thermal processing of HNCO the most favorable production mechanism for OCN$^{-}$.\\   
%fig13
\section{Conclusion}
We have presented a laboratory study on the efficiency of thermal versus UV-mediated OCN$^-$ formation to qualitatively investigate the interstellar 4.62\,$\mu$m feature. OCN$^{-}$ is easily formed by UV-photolysis of a classic ice matrix of H$_{2}$O/CO/NH$_{3}$ and also when CO is replaced by CH$_{3}$OH or HNCO. Nevertheless, it remains questionable whether sufficient UV-photons are present inside dense clouds. UV photolysis also requires high initial abundances of NH$_{3}$ to reproduce the optical depth of the 4.62\,$\mu$m feature, particularly towards YSO's where observed OCN$^{-}$ abundances exceed 1.2\% with respect to H$_{2}$O. Moreover, when starting from a CO ice mixture, the amount of NH$_{3}$ remaining in the ice after photoprocessing is too high to be consistent with the conservative upper limit of 5\% determined toward W\,33\,A. In a CH$_{3}$OH ice mixture a lower NH$_{3}$ content is needed but the UV fluence required is approximately 3$\times$ higher with respect to a CO-containing ice. The most favorable OCN$^-$ formation route emerging from this study is the thermal processing of HNCO-containing ices. Provided that HNCO is present in interstellar ices, thermal processing is $\sim$100\% efficient and is consistent with the data in particular toward W\,33\,A, NGC 7538 IRS9 and GL 2136. Also the inferred interstellar ice temperatures along the line-of-sight towards other OCN$^-$-containing YSO's are in agreement with this route of formation.\\
In the case of sources for which OCN$^-$ has not been detected, little insight can be gained about its formation mechanism. They could be either too young (too little UV exposure) or colder than 10 K. In these cases, HNCO can be present but only in H$_2$O-rich ices at temperatures below 10 K. At higher temperatures, the presence of sufficient amounts of proton acceptors like NH$_3$ rapidly induces OCN$^-$ formation. A non-detection for OCN$^-$ would in that case indicate lack of HNCO or NH$_3$ formation. Future searches for weak OCN$^-$, NH$_3$ and HNCO features are needed to further constrain these chemical routes.
\begin{acknowledgements}
The authors wish to thank Helen Fraser and Klaus Pontoppidan for interesting discussions and Ewine van Dishoeck for many helpful comments on the paper. This research was financially supported by the Netherlands Research School for Astronomy (NOVA) and a NWO Spinoza grant.  
\end{acknowledgements}
{}
\end{document}